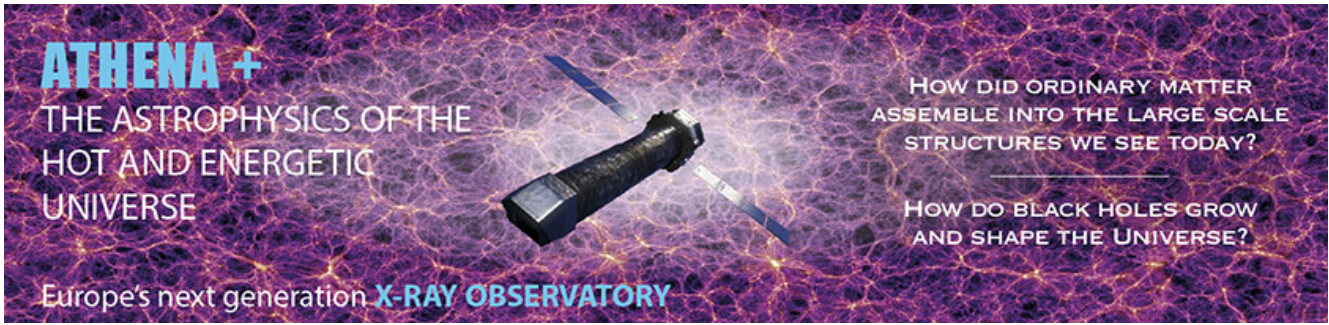

# The Hot and Energetic Universe

An *Athena+* supporting paper

## The Wide Field Imager (WFI) for Athena+

## Authors and contributors

**A. Rau,** N. Meidinger, K. Nandra, M. Porro, D. Barret, A. Santangelo, C. Schmid, L. Strüder, C. Tenzer, J. Wilms, C. Amoros, R. Andritschke, F. Aschauer, A. Bähr, B. Günther, M. Fürmetz, B. Ott, E. Perinati, D. Rambaud, J. Reiffers, J. Treis, A. von Kienlin, G. Weidenspointner

# 1. EXECUTIVE SUMMARY

The Wide Field Imager (WFI) is one of the two scientific instruments proposed for the *Athena+* X-ray observatory. It will provide imaging in the 0.1-15 keV band over a wide field, simultaneously with spectrally and time-resolved photon counting. The instrument is designed to make optimal use of the grasp (collecting area times solid angle product) provided by the optical design of the *Athena+* mirror system (Willingale et al. 2013), by combining a sensitive approx. 40' diameter field of view (baseline; 50' goal) DEPFET detector with a pixel size properly sampling the angular resolution of 5" on-axis (half energy width). This synthesis makes the WFI a very powerful survey instrument, significantly surpassing currently existing capabilities (Nandra et al. 2013; Aird et al. 2013). In addition, the WFI will provide unprecedented simultaneous high-time resolution and high count rate capabilities for the observation of bright sources with low pile-up and high efficiency. In this paper, we summarize the instrument design, the status of the technology development, and the baseline performance.

# 2. INSTRUMENT DESIGN

## 2.1. Detector Concept

The technology and design of the WFI for *Athena+* builds on the strong heritage of the wide field imagers proposed for the International X-ray Observatory and for Athena (e.g., Stefanescu et al. 2010). The heart of the camera is formed by a set of arrays of DEPFET (DEpleted P-channel Field Effect Transistor; Kemmer & Lutz 1987) active pixels integrated onto common 450 µm thick silicon bulks. Similar sensors have been developed for the Mercury Imaging X-ray Spectrometer (MIXS) for ESA's *BepiColombo* mission to Mercury to be launched in 2015 (e.g., Treis et al. 2008) and will also be used in a variety of ground-based experiments (e.g., European X-ray Free Electron Laser; Porro et al. 2008).

A DEPFET is a combined detector-amplifier structure (Figure 1, left). Every pixel consists of a p-channel Metal-Oxide Semiconductor Field Effect Transistor (MOSFET), which is integrated onto a fully (sideways) depleted silicon bulk. With an additional deep-n implantation, a potential minimum for electrons, the so-called internal gate, is generated and laterally constrained to the region below the transistor channel. Incident X-ray photons interact with the bulk material, and generate a number of electron hole pairs proportional to the incident photon energy. Holes are removed over the backside contact while electrons are collected in the internal gates of the pixels nearest to the photon interaction site. The conductivity of the MOSFET channel will be modulated by their presence. The change in conductivity is proportional to the number of charge carriers and therefore a measure of the energy of the incident photon. The internal gate and the nearby clear gate and clear contact form an n-channel MOSFET, enabling the removal of collected charges by applying sufficient voltages.

The internal gate persists regardless of the presence of a transistor current. Thus, each row of pixels be turned off during a certain exposure time, and only turned on for readout on demand. The amount of integrated charge can then be sensed by turning on the transistor current and measuring the conductivity before and after the charge removal (Figure 1, right). The difference in conductivity is then proportional to the amount of charge collected in the internal gate.

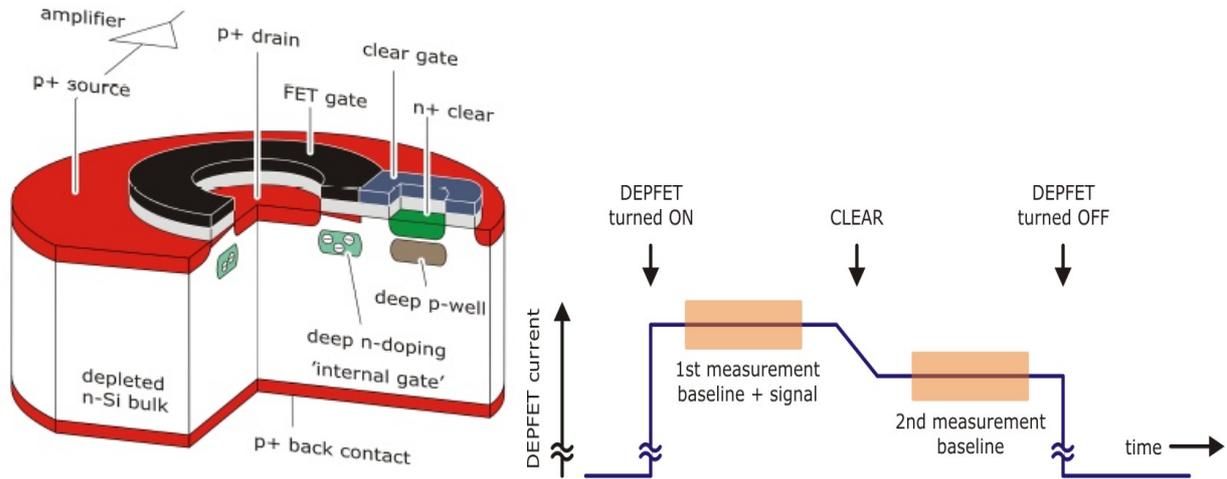

**Figure 1:** *Left:* Cutaway display of a circular MOS-type DEPFET. *Right:* Signal evaluation scheme for the readout of an active pixel sensor. After the device is turned on, the signal level (current or voltage) is determined before and after the clear. The difference signal is then proportional to the removed charge during the clear.

In normal operation mode, photons hitting the sensor during the moments of integration or clear will receive an incorrect energy association. This leads to an increased background level and a spectral distortion. In order to suppress these events, a shutter can be implemented into each active pixel. Two layout concepts are currently studied (Lutz et al. 2007, Bähr et al. 2012). Both include an additional highly n-doped contact serving as sink for electrons produced during read out and a potential-barrier to avoid electrons from the internal gate. These so-called gateable DEPFETs have all the benefits of the normal sensors and at the same time provide a fast built in shutter. Using recently produced prototypes a shielding of the collection anode up to $10^{-4}$ was achieved. Simulations predict that values of $10^{-5}$ and better can be reached with an optimized geometry. In addition to the shutter capability, pixel layouts containing an intermediate storage region are in development. Here, generated during the readout, is not lost but accumulated and preserved for later processing. This capability completely obviates dead times and the distortion of the spectral resolution while maximizing the throughput (Figure 2).

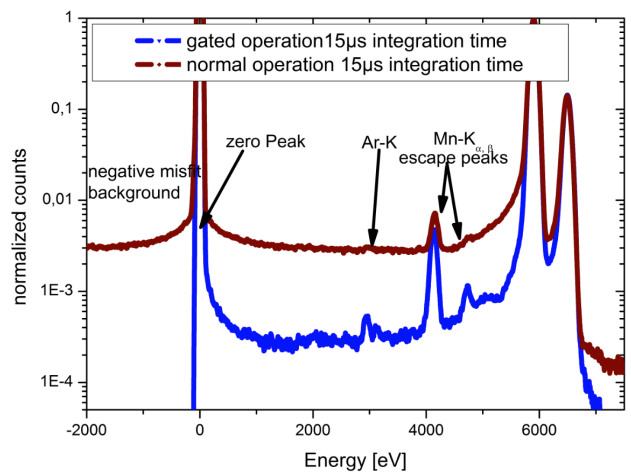

**Figure 2:** Impact of the gateable DEPFET technology demonstrated by $^{55}$Fe spectra measured with a macro-pixel in gated (blue) and normal (dark red) operation mode. Without shutter, charges deposited during the integrations will cause misfits. If happening during the first integration, a lower energy is derived, leading to a higher background signal. Charges during the second integration lead to negative differences and produce background at apparent negative energies.

Besides the capability of completely negating out-of-time events, the DEPFET technology provides several advantages to standard CCD based detectors. E.g., the signal charges are not shifted, charge transfer losses do not occur. Through this, radiation damages that have negative effects on the CTE of CCD's have lower impact on DEPFET detectors, ensuring a intrinsically higher radiation hardness.

## 3.2. Focal Plane Design

The WFI will combine in a single focal plane array its excellent wide field survey power with its high-count rate and timing capabilities. Due to the physical size, this cannot be realized with a single chip on a monolithic wafer. Instead, the current design foresees a mosaic of multiple sensor chips, which together fill the field of view provided by the Athena+ mirror system.

Several different layout options are currently studied. One possibility for the Athena+ mirror system baseline diameter of 40' is depicted in Figure 3. Here, the center of the mosaic is formed by a 256 x 256 matrix of 100x100µm² pixels covering a field of view of approx. 450" x 450". The selected pixel size properly samples (factor 2.8) the 5" on-axis PSF and at the same time retains the high count rate capability (see also Section 3.4). This central chip is surrounded by four 448 x 640 matrices of 130x130µm² pixels, each covering approx. 1022" x 1460". The slightly larger pixel size will sufficiently sample the PSF (factor 2.2-3.4 depending on the off-axis angle). All together, a square field of view of ~40' x 40', would be filled with active pixels.

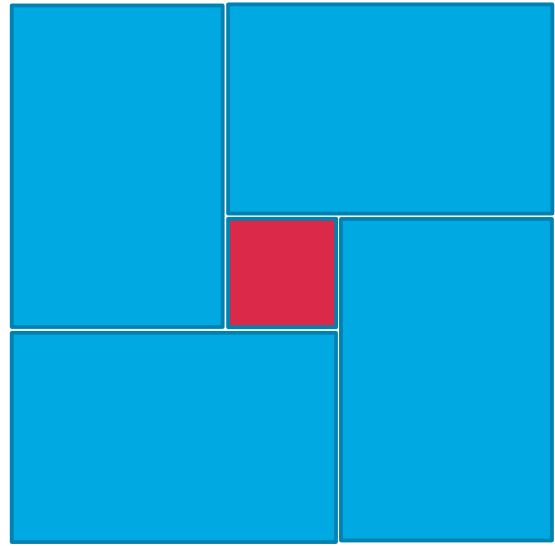

**Figure 3:** Layout of the DEPFET sensor arrays in the focal plane. The positions of the central (shown in red) and outer (blue) chips are indicated schematically.

The layout can easily be modified to support the larger (50') goal FoV for the Athena+ mirror system by increasing the number of pixels of the outer DEPFET matrices (e.g., to 576 x 768 per chip).

## 2.3. Front End Electronics

The WFI focal plane array requires two different frontend Application-Specific Integrated Circuit (ASIC) devices, the control front end (CFE) and the analog front end (AFE). Pixels are controlled by toggling a sequence of voltages on the gate, clear gate, and clear contacts of each row, and sensing the current through each column. Here, the correct voltage sequence is applied by the SWITCHER CFE, shown exemplary in Figure 4 (left and right of the active pixel sensor array). For the AFE we currently develop a new low-noise multi-channel signal amplifier/shaper circuit with integrated sequencer and serial analog output (VERITAS2; Porro et al. 2013). VERITAS2 is based on a new circuit topology and implements a trapezoidal filter function with a single fully-differential amplifier. The trapezoidal filter function is the time-limited optimum filter for white series noise, which is dominant at the foreseen readout speed.

The most important feature established with VERITAS2 is, additional to the source-follower, the drain-readout. This is made possible by a low-noise current-to-voltage converter placed in front of the preamplifier. The main advantage of the drain-readout is that all the nodes of the DEPFET are at a fixed potential and the readout speed of the system is not limited by the resistor-capacitator (RC) time constant of the input of the readout chain. This capacitance comprises the input capacitance of the preamplifier and the parasitic capacitance associated to the DEPFET source line. Coupling the ASIC in the drain-readout mode with a DEPFET array, it will be possible to obtain a readout time per row of about 2 µs. The outputs of the analog channels of VERITAS2 are serialized by a

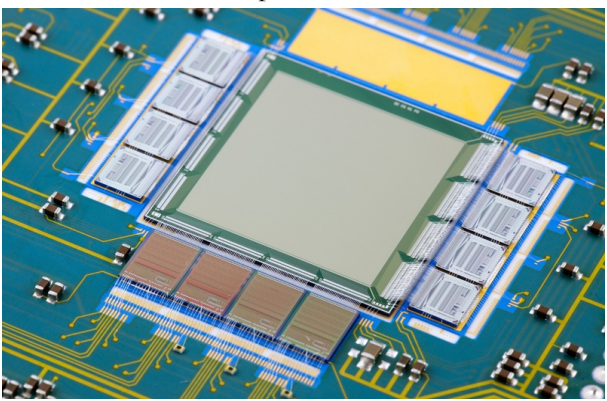

64:1 multiplexer with a clocking speed up to 20 MHz and sent to a fast fully differential output buffer. The architecture allows window-mode readout of the pixel matrices making it possible to address selectively arbitrary sub-areas of the DEPFET matrix or even to readout different sub-areas at different speeds.

**Figure 4:** DEPFET hybrid with a prototype 256 x 256 sensor chip in the center, four banks of CFE SWITCHER control chips left and right, and four AFE ASIC readout chips at the bottom. (photo from Meuris et al. 2011)

To minimize the gap between the sensor arrays, and thus maximize the field of view covered with active pixels, the CFE and AFE of the central chip cannot be placed as shown in Figure 4 but will be folded to the backside of the detector. Detailed design studies are under way.

## 2.4. Filters and Quantum Efficiency

The sideways depletion utilized for the DEPFETs provides for an unobstructed, homogenous entrance window with a 100 % filling factor and excellent quantum efficiency (QE) at low energies (Figure 5, left). This low energy sensitivity is conserved by the use of a 70nm thin $SiO_2/Si_3N_4$ entrance window. The 450 μm thickness of the depleted Si bulk provides also a high QE above 10keV, thus the overall accessible energy range comprises approx. 0.1-15 keV. However, the large effective area Athena+ X-ray mirror system will focus in addition to the X-ray photons also a significant amount of optical and UV light on to the detector. Most of these photons will be blocked by the multi-layer entrance window located directly on top of the active pixel sensors and by an additional thin (e.g., 70 nm) Al layer deposited on top of that. For the remaining optical/UV photons two alternative approaches are currently studied. The thickness of the on-chip filter could be chosen such that even for optically bright sources optical loading is sufficiently reduced. Alternatively, a thin 40 nm Al-layer on top of a 350 nm polymer carrier in a filter wheel could be used similar to the concept developed for the WFI for *Athena* (Figure 5, right). The latter could rotate in and out of the photon path to facilitate observations where the low-energy response is scientifically critical. However, the physical size of the field of view places strong requirements on the stability and longevity of such a thin filter.

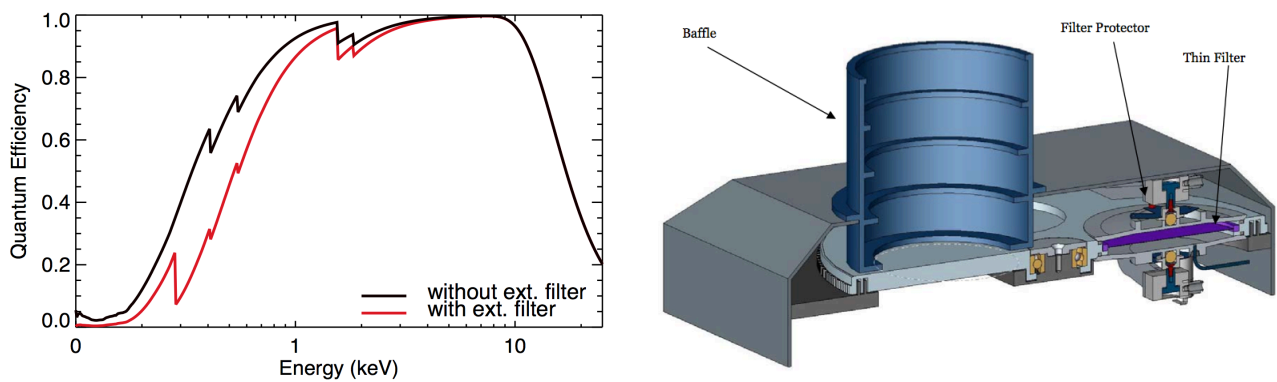

**Figure 5:** *Left:* Quantum efficiency of the DEPFET sensor as function of energy without (black curve) and with (red curve) external blocking filter for optical/UV light. *Right:* Filter wheel concept with X-ray baffle as designed for the *Athena* mission proposal. In order to protect the thin and large filter against the acoustic noise during the rocket launch a filter protector is necessary. In comparison to the design proposed for *Athena*, the physical diameter of the *Athena+* WFI has nearly doubled, requiring a significantly larger filter membrane and protector.

## 2.5. Architecture

Before the information about a detected X-ray photon can reach the astronomer, a number of processing steps are required. A summary is presented in Figure 6 and a concept design of the instrument architecture shown in Figure 7. After passing through the X-ray mirror system, a photon is collected in the active pixel array and subsequently converted into a charge. After a first amplification in the DEPFET itself the signal is further amplified and shaped in the AFE VERITAS2 and there converted into an analog voltage signal. The DEPFET arrays and front end electronics are mounted on the camera head, which provides the structural stability and required cooling resources. The signal is then transferred to the pre-processor using a short flexlead. The pre-processor contains the analog-digital-converters (ADCs) digitizing the output from the VERITAS2. This digitization takes places in so-called ADC clusters, cards hosting multiple ADCs and one field-programmable gate array (FPGA). Furthermore, the pre-processor performs the necessary offset calculation and subtraction, noise and threshold calculation, as well as gain and common mode correction. The frame builder collects the data from the pre-processor ADCs and performs pattern recognition, event identification and data compression, and prepares the data for transmission to the ground.

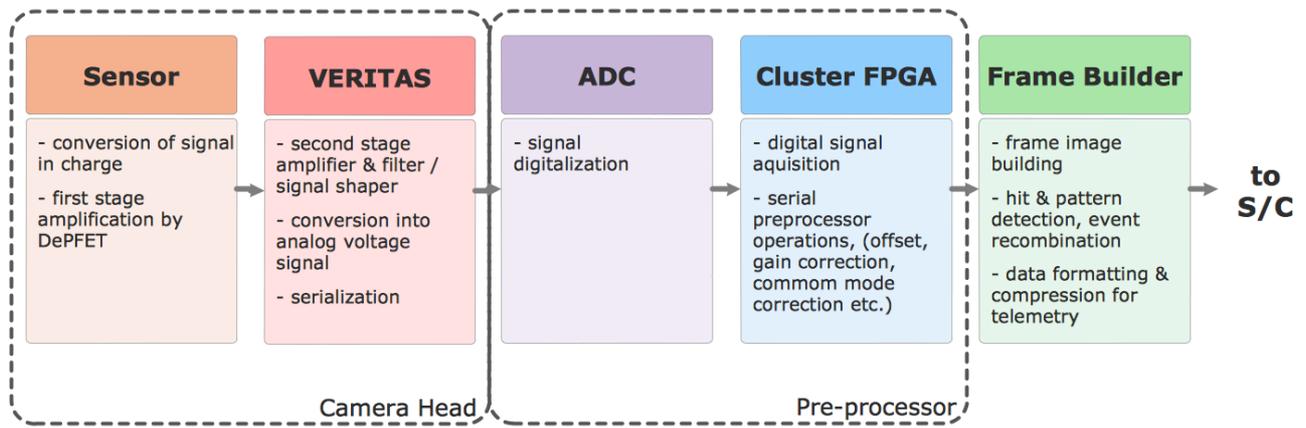

**Figure 6:** Overview of the various data processing steps from the sensing of the incoming photons to the telemetry transmission to the spacecraft.

The electronics responsible for controlling the instrument is located in the instrument control unit (ICU) box. The ICU holds beside the frame builder (see above), also the instrument processor, power conditioner and mechanism controls. The instrument processor is the main computer controlling the WFI and performs all data interface tasks with the spacecraft and commands all aspects of the WFI. The power conditioner is responsible for generating the voltages required for the operation of the camera head and the electronic boxes from the power provided by the spacecraft. The mechanism controller is in charge of controlling the filter wheel and the solid state relays.

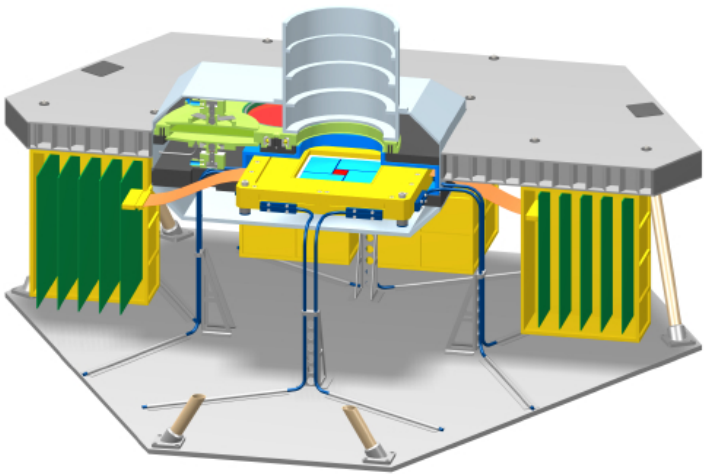

**Figure 7:** Cut through the concept drawing of the WFI. At its heart lies the camera head hosting the sensors and front end electronics. On top of the camera head lies the X-ray baffle and filter wheel (see also Section 2.4). Flexleads connect to the electronic boxes holding the pre-processors. Heat pipes transporting the excess heat from the detector are shown in blue.

## 3. INSTRUMENT PERFORMANCE

### 3.1. Effective Area

The WFI on-axis response results from the combination of the effective area of the telescope (see Willingale et al. 2013), the attenuation by the UV/optical light blocking filter, and the quantum efficiency of the detector (see Section 2.4). Figure 8 (left) shows the impact of the latter two components by comparing the input telescope response with the effective area as sensed by the detector on-axis. While the effective area remains practically unaffected above 1 keV and reaches approx. 1.7 m$^2$ at 1 keV, an increasing reduction with decreasing photon energy compared to the effective area provided by the mirror system is immanent.

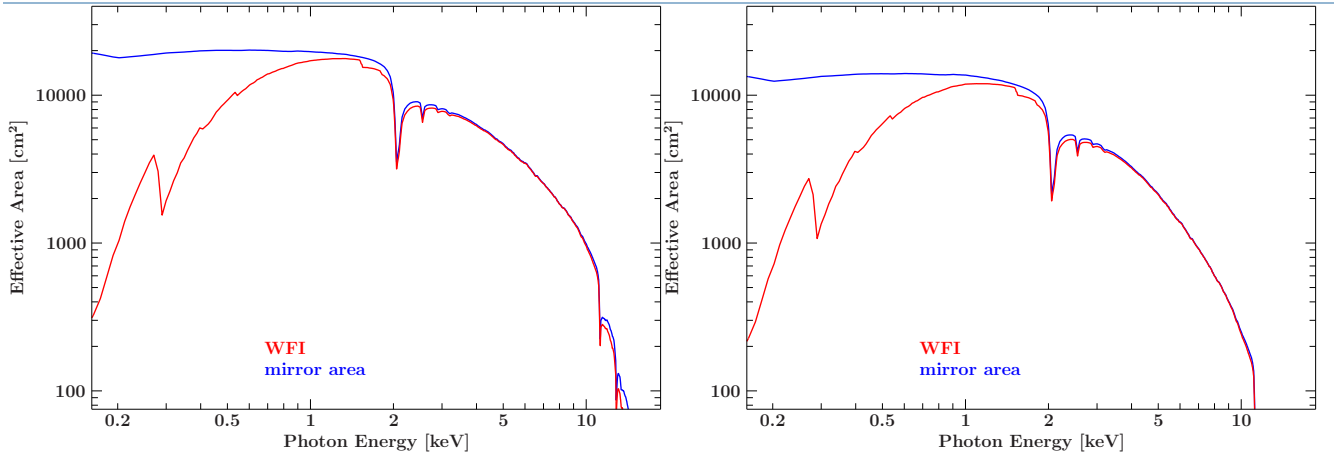

**Figure 8:** *Left*: On-axis response of the WFI (red) compared to the on-axis effective area provided by the *Athena+* mirror system (blue). Apart from the mirrors, the instrumental effective area also takes into account the quantum efficiency of the detector and the attenuation by the optical/UV light blocking filter. *Right*: Same as left but including the field-of-view-averaged vignetting.

The collimation imposed by the pore geometry of the Silicon Pore Optics mirror modules leads to a loss in effective area towards the edge of the field of view (see Willingale et al. 2013). This vignetting depends strongly on the photon energy and is strongest at higher photon energies. The WFI response function taking into account the field-of-view averaged vignetting is presented in Figure 8 (right) and reflects the overall reduced effective area compared to the on-axis behavior.

### 3.2. Spectral Resolution

The spectral resolution of the DEPFET sensors is primarily limited by the statistical variation in the charge generated by the interacting X-ray photon. This so-called Fano-limited energy resolution (Fano 1947) is reached even at fast readout speeds. The performance has been verified with laboratory measurements by irradiating a prototype DEPFET sensor matrix with photons from X-ray calibration sources. Figure 9 shows the resulting spectrum of an $^{55}$Fe source with its prominent Mn K$\alpha$ (5.9 keV) and Mn K$\beta$ (6.5 keV) lines. The resolution at 5.9 keV is 125 eV (FWHM) in agreement with the expectation from the Fano limit and the measurement shows a good peak-to-valley ratio of approx. 2900. Also, the $C_K$ line at 277 eV can still be resolved with an energy resolution of 50 eV. The detection efficiency at low energies is mainly determined by the quality of the radiation entrance window and the integrated optical/UV light blocking filter.

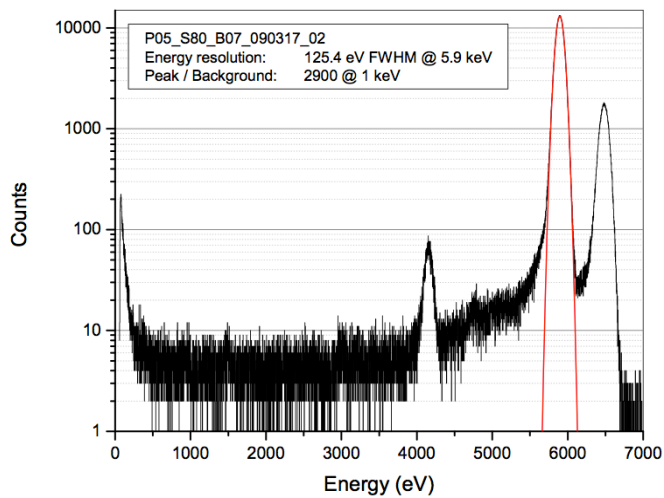

**Figure 9**: Backside-illuminated $^{55}$Fe X-ray spectrum taken with a DEPFET matrix. In addition to the 5.9 keV and 6.5 keV Mn K$\alpha$ and Mn K$\beta$ lines, also the Si escape peak at 4.2 keV can clearly be detected.

The spectral resolution of the DEPFET detectors is expected to remain constant throughout the mission lifetime. In contrast, the energy resolution of CCDs can degrade over time due to an increased charge transfer inefficiency (CTI) caused by encounters with low energetic protons. The latter had, e.g., a significant impact on the performance of the Chandra ACIS-I detectors, degrading them from 130 eV (at 5.9 keV) at launch to a row-dependent resolution of up to ~300 eV in 2012. The Athena+/WFI will therefore not only provide the advantage of a significantly larger effective area (and thus more photons in the deep fields) but also a more than two times better energy resolution compared to the current ACIS-I performance.

## 3.3. Background

The sensitivity of the WFI, in particular for faint sources, depends strongly on the incident background photon flux. Two main components contribute to this background, the diffuse cosmic X-ray background and secondary X-rays produced in the interaction of charged particles with the spacecraft and instrument material. The cosmic photon background consists of the integrated emission from unresolved extragalactic point sources and two main diffuse Galactic foreground components, one associated with the hot diffuse gas in the local bubble and a second arising from hot diffuse gas in the Galactic halo.

The cosmic X-ray background flux per unit area is solely defined by the effective area of the *Athena+* optics system and cannot be shielded inside the camera without at the same time reducing also the photon flux from the astrophysical target of interest. However, the secondary fluorescence photons induced by the particle background can be rejected with high efficiency. For this purpose, the WFI detectors will utilize a graded-Z shield, consisting of a laminate of several materials involving a gradient from high Z (e.g., Tantalum) to low Z values (e.g., Carbon). The gradient is selected such that subsequent layers absorb fluorescence emission from the outer layers and thus effectively suppress any fluorescent emission lines. The remaining background is essentially a continuum dominated by secondary electrons produced in the graded-Z shield itself together with a factor of ten smaller contributions from alpha particles, positrons, and protons.

The performance has been verified with GEANT4 simulations for the *Athena*/WFI concept in an L2 orbit (Hauf et al. 2011). With the exception of the Si K$\alpha$+$\beta$ lines at 1.74 keV and 1.84 keV produced in the sensor itself, all other fluorescent emission lines are quenched and the background can be described by a power law continuum with a slope of 0.15 and a normalization of $1.4 \times 10^{-4}$ cnt/keV/s/amin$^2$ at 1 keV. These simulations have been performed for solar activity minimum, i.e., cosmic ray maximum, and thus already provided a conservative estimate. However, the evolution of the camera design from *Athena* to *Athena+* still requires the new simulations to be performed. Until then, a conservative increase of the normalization of the particle background by a factor 2 over the above described simulation for the entire field of view can be assumed.

The combined cosmic X-ray background and particle induced background is shown in Figure 10. The particle background becomes significant only above approx. 2 keV.

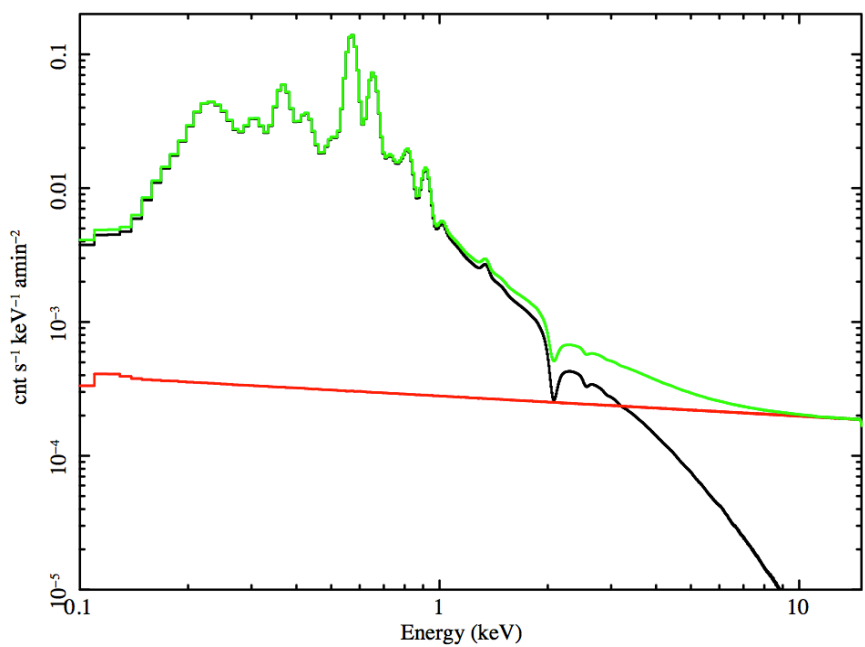

**Figure 10:** Predicted photon background (green) for the WFI composed of contributions from the cosmic X-ray background (black) and from the particle interactions (red).

### 3.4. Readout Speed and HTR capabilities

To achieve the required high-speed performance, the central 256x256 pixel DEPFET detector (see Section 2.2) will be subdivided into two hemispheres, read out in parallel by four readout ASICs per hemisphere. Although each ASIC has a readout time per row of about 2 µs, the effective signal processing time per row can further be halved to 1 µs by independently reading out two rows per hemisphere simultaneously. This allows to read out the full chip in 128 µs, corresponding to a rate of approx. 7.8 kHz. The expected percentages of recognized valid patterns and of pile-up as function of incident photon rate for the baseline on-axis PSF of 5" and the goal PSF of 3" are shown by the red data in Figure 11.

Reducing the number of readout rows further optimizes the achievable time resolution. For a point source on-axis, a split frame readout can be used, which reads eight rows per hemisphere in parallel. These 16 rows comprise a width of approx. six times that of the 5" PSF and can be read out in a total of 8 µs. The resulting dependencies are shown with the green data points in Figure 11. For the baseline PSF of 5" a source with a brightness of 0.5 Crab (1.0 Crab) will see a throughput of >88 % (>80 %) and suffer a pile-up of only <3 % (<5 %).

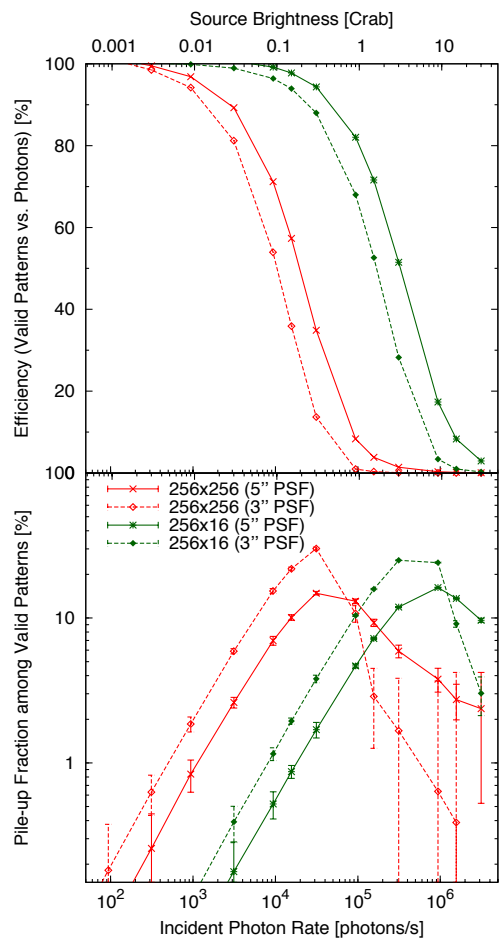

**Figure 11:** *Top:* Simulated detector efficiency as function of incident photon rate in units of ph/s and Crab for two different windows (256 x 256, 256 x 16 pixels) and two on-axis PSFs (3", 5"). *Bottom:* Pile-up fraction as function of incident photon rate.

## 4. INSTRUMENT CHARACTERISTICS

A summary of the key instrument characteristics is presented in Table 1.

The Wide Field Imager has been designed to enable a significant part of the highly demanding science aims of the *Athena+* mission. Complementary to the second instrument on board the observatory, the high-spectral resolution the X-ray Integral Field Unit(X-IFU), the WFI provides the important wide field survey capability with a high sensitivity, a moderate spectral resolution, and excellent count rate capacity. It will make optimal use of the factor of 10 larger throughput of the *Athena+* optics system compared to XMM Newton over a field of view that is a factor of >1.8 larger than that of XMM Newton and a factor of 5 larger than that of Chandra. Taking also into account the significantly better field-of-view-averaged angular resolution compared to XMM, the WFI will provide a dramatic leap in survey sensitivity to any existing X-ray mission. More details on how the WFI will impact the various science areas associated with the *Athena+* theme "The Hot and Energetic Universe" can be found in the white paper submitted to ESA (Nandra et al. 2013) and in the reports prepared by the science working groups available at http://www.the-athena-x-ray-observatory.eu.

**Table 1:** Instrument characteristics.

| Parameter | Characteristic |
|---|---|
| **Energy Range** | 0.1-15 keV |
| **Field of View** | ca. 40' x 40'  (baseline) |
| | ca. 50' x 50' (goal) |
| **Array Format** | Central chip: 256 x 256 pixel |
| | Outer chips: 4x 448 x 640 pixel (baseline) |
| | 4x 576 x 768 pixel (goal) |
| **Pixel Size** | Central chip: 100 x 100 µm² (1.8") |
| | Outer chips: 130 x 130 µm² (2.3") |
| **Angular Resolution (onaxis)** | <5 arcsec (oversampling by 2.8) |
| **Quantum efficiency (incl. optical blocking filter)** | 277 eV: 24% |
| | 1 keV: 87% |
| | 10 keV: 96% |
| **Energy Resolution** | ΔE < 150 eV (FWHM) @ 6 keV |
| **Readout rate** | Central chip: 7800 fps |
| | Outer chips: 2200 fps |
| **Fast timing, count rate capability** | 8 µs in window mode |
| | 0.5 Crab > 88 % throughput, <3 % pile-up |
| | 1 Crab > 80 % throughput, <5 % pile-up |
| **Particle Background at L2** | 3 x 10$^{-4}$ cnt/cm²/keV/s |

## 5. REFERENCES

Aird, J., et al., 2013, arXiv:1306.2325
Bähr, A., et al., 2012, SPIE, 8453, 13
Fano, U., 1947, Phys. Rev., 72, 26
Hauf, S. et al. , 2011, SPIE, 8443, 5
Kemmer, J., Lutz, G., 1987, Nucl. Instr. And Meth A, 253, 356
Lutz, G., et al., 2007, IEEE Nucl. Science Symp. Conf. Rec, 2, 988
McCammon, D., et al., 2002, ApJ, 576, 188
Meuris, A., et al., 2011, IEEE Trans. On Nucl. Science, 58, 1206
Nandra, P., et al., 2013, arXiv:1306.2307
Porro, M., et al., 2008, in Proc of IEEE Nucl. Science Symp. Conf. Rec., N14-7, 1578
Porro, M., et al., 2013, IEEE Trans. On Nucl. Science, 60, 446
Stefanescu, A. et al. 2010, Nucl. Instr. And Meth A, 624, 533
Treis, J., et al. 2008, SPIE, 7021, 29
Willingale R., et al., 2013, arXiv:1307.1709